\documentclass[pre,twocolumn,showpacs,superscriptaddress]{revtex4}

\usepackage{graphicx}
\usepackage{amsmath}
\usepackage{amssymb}
\usepackage[british]{babel}

\begin{document}

\title{Kramers escape driven by fractional Brownian motion}

\author{Oleksii Yu. Sliusarenko}
\email{aslusarenko@kipt.kharkov.ua}
\affiliation{Akhiezer Institute for Theoretical Physics NSC KIPT,
Akademicheskaya Str.1, 61108 Kharkov, Ukraine}
\author{Vsevolod Yu. Gonchar}
\email{vsevolod.gonchar@gmail.com}
\affiliation{Akhiezer Institute for Theoretical Physics NSC KIPT,
Akademicheskaya Str.1, 61108 Kharkov, Ukraine}
\author{Aleksei V. Chechkin}
\email{achechkin@kipt.kharkov.ua}
\affiliation{Akhiezer Institute for Theoretical Physics NSC KIPT,
Akademicheskaya Str.1, 61108 Kharkov, Ukraine}
\affiliation{School of Chemistry, Tel Aviv University, Ramat
Aviv, Tel Aviv 69978, Israel}
\author{Igor M. Sokolov}
\email{igor.sokolov@physik.hu-berlin.de}
\affiliation{Institut f{\"u}r Physik, Humboldt Universit{\"a}t
zu Berlin, Newtonstra{\ss}e 15, 12489 Berlin, Germany}
\author{Ralf Metzler}
\email{metz@ph.tum.de}
\affiliation{Physik Department, Technical University of Munich,
James Franck Strasse, 85747 Garching, Germany}

\begin{abstract}
We investigate the Kramers escape from a potential well of a test particle
driven by fractional Gaussian noise with Hurst exponent $0<H<1$. From a
numerical analysis we demonstrate the exponential distribution of escape
times from
the well and analyze in detail the dependence of the mean escape time as
function of $H$ and the particle diffusivity $D$. We observe different
behavior for the subdiffusive (antipersistent) and superdiffusive
(persistent) domains. In particular we find that the escape becomes
increasingly faster for decreasing values of $H$, consistent with previous
findings on the first passage behavior. Approximate analytical calculations
are shown to support the numerically observed dependencies.
\end{abstract}


\maketitle

\section{Introduction}
\label{sec:introduction}

Anomalous diffusion is characterized by a deviation from the classical
linear time dependence $\langle x^2(t)\rangle\simeq t$ of the mean squared
displacement. Such anomalies range from ultraslow transport $\langle
x^2\rangle\simeq\log^{\beta}t$ as discovered in Sinai diffusion or in
iterated maps \cite{sinai,julia}, up to cubic diffusion $\langle x(t)\rangle
\simeq t^3$ in random walk processes with correlated jump lengths
\cite{vincent1} or the relative coordinate of two particles encountered in
turbulent Richardson flow \cite{richardson,boffetta}. Here
we are interested in anomalous diffusion of the power-law type
\cite{bouchaud,report}
\begin{equation}
\label{msd}
\langle x^2(t)\rangle=2Dt^{2H},
\end{equation}
where $H$ is the Hurst exponent and $D$ the generalized diffusion
coefficient of dimension $[D]=\mathrm{cm}^2/\mathrm{sec}^{2H}$.
Depending on the magnitude of $H$ we observe subdiffusion
($0<H<1/2$) or superdiffusion ($1/2<H<1$). The limits $H=1/2$ and
$H=1$ correspond to ordinary Brownian diffusion or ballistic motion,
respectively. For one-particle motion ballistic transport is the upper
limit of spreading when the particle has a finite maximum velocity.

Anomalous diffusion of the power law form (\ref{msd}) is observed in a
multitude of systems. In particular, subdiffusion was found for the motion
of charge carriers in amorphous semiconductors \cite{scher,pfister}, the
spreading of tracer molecules in subsurface hydrology \cite{scher_grl},
diffusion on random site percolation clusters \cite{klemm} as well as the
motion of tracers in the crowded environment of biological cells \cite{bio}
or in reconstituted biological systems \cite{alabio},
among many others. Examples for superdiffusion include active motion in
biological cells \cite{caspi}, tracer spreading in layered velocity fields
\cite{matheron}, turbulent rotating flows \cite{swinney}, or in bulk
mediated surface exchange \cite{stapf}.

Apart from numerical approaches there exist two prominent analytical models
for such
anomalous diffusion: One is the continuous time random walk (CTRW) model
\cite{scher,klablushle} in which each jump is characterized by a variable
jump length and waiting time drawn from associated probability densities.
CTRW theory includes (i) subdiffusion when the variance of jump lengths
is finite but the waiting times have an infinite characteristic time; (ii)
L{\'e}vy flights when the mean waiting time is finite but the jump length
variance diverges;
and (iii) L{\'e}vy walks in which waiting times and jump lengths are coupled,
producing sub-ballistic superdiffusion with finite variance. The escape over
a potential barrier for subdiffusion and L{\'e}vy flights was studied recently
\cite{mekla,levykramers,imkeller,ditlevsen}.

The second model is fractional Brownian motion (FBM). It was
originally described by Kolmogorov \cite{kolmogorov} and
reintroduced by Mandelbrot and van Ness \cite{mandelbrot}. FBM is a
self-similar Gaussian process with stationary increments
\cite{Yaglom,Qian}. The FBM mean squared displacement follows
Eq.~(\ref{msd}), and the Hurst exponent $H$ of the fractional
Gaussian noise varies in the full range $0<H<1$. Uncorrelated,
regular Brownian motion corresponds to $H=1/2$. For $0<H<1/2$ the
prefactor in the noise autocorrelation is negative, rendering the
associated antipersistent process subdiffusive. That means that a
step in one direction is likely followed by a step in the other
direction. Conversely, in the case $1/2<H<1$ the motion is
persistent, effecting sub-ballistic superdiffusion in
which successive steps tend to point in the same direction.
FBM is used to model a variety of processes including
monomer diffusion in a polymer chain \cite{gleb}, single file
diffusion \cite{tobias}, diffusion of biopolymers in the crowded
environment inside biological cells \cite{guigas}, long term storage
capacity in reservoirs \cite{hurst}, climate fluctuations
\cite{palmer}, econophysics \cite{mandelbrot1}, and teletraffic
\cite{mikosch}.

Despite its wide use FBM is not completely understood. Thus the general
incorporation of non-trivial boundary conditions is unattained, in particular,
the first passage behavior is solved analytically solely on a semi-infinite
domain \cite{Molchan}. Notably the method of images does not apply to solve
boundary value problems for FBM. Similarly the associated fractional Langevin
equation driven by fractional Gaussian noise was recently discovered to exhibit
critical dynamical behavior \cite{stas}.

\begin{figure}
\includegraphics[width=8.8cm]{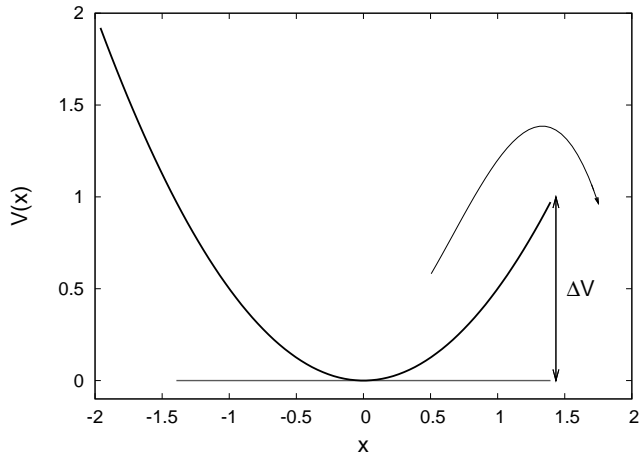}
\caption{Harmonic potential $V(x)=x^2/2$ with cutoff at $x=\sqrt{2}$ used
in the numerical analysis of the FBM Kramers escape. The potential barrier
height is $\Delta V=1$. Dimensionless units.}
\label{potential}
\end{figure}

Here we study the generalization for FBM of the Kramers escape from a potential
well across a finite barrier, as illustrated in Fig.~\ref{potential}. This
problem is relevant, for instance, for single file diffusion in external
potentials \cite{eli}, the dissociation dynamics
of biopolymers from a bound state in FBM models for particle diffusion under
molecular crowding conditions \cite{guigas} or bulk chemical reactions of
larger particles under superdense conditions. We note that a similar problem
was treated for correlated Gaussian noise \cite{katja} and for fractional
Langevin equation motion in the case when the fluctuation dissipation theorem
applies \cite{goychuk}. We here study the important case of external
fluctuations, that is, for systems which do not obey the fluctuation
dissipation theorem \cite{klimontovich}.

In the regular Kramers theory \cite{kramers,chandrasekhar,risken} for the
escape of a Brownian particle across a potential barrier in the high barrier
limit $\Delta V\gg k_B\theta$, where $k_B\theta$ denotes thermal energy, the
probability density of the first escape from the well follows an exponential
decay,
\begin{equation}
\label{decay}
p(t)\simeq\exp\left(-\frac{t}{T}\right).
\end{equation}
This corresponds to the relaxation mode of the lowest eigenvalue
\cite{kramers,chandrasekhar,risken}. In Eq.~(\ref{decay})
the characteristic escape time $T$ is proportional to the Arrhenius
factor of the barrier height $\Delta V$,
\begin{equation}
\label{arrh} T\propto\exp\left(\frac{\Delta V}{k_B\theta}\right).
\end{equation}
In what follows we demonstrate from simulations and analytical considerations
that the exponential decay (\ref{decay}) is preserved in FBM processes due to
the stationary nature of FBM, while the activation pattern (\ref{arrh})
becomes explicitly dependent on the Hurst exponent. This $H$-dependence is
different for the antipersistent and persistent cases. Remarkably slow
diffusion leads to fast escape, that is, the lower the value of $H$ is
chosen the faster the escape from the potential well becomes. This observation
is consistent with the first passage behavior of FBM that is known analytically,
and analysed numerically in the Appendix.

We first investigate FBM driven Kramers escape by numerical integration of
the Langevin equation subject to fractional Gaussian noise in Sec.~\ref{sim}.
In particular, we analyze the distribution of escape times and the dependence
of the mean escape time on the Hurst exponent $H$ and the noise strength $D$.
In Sec.~\ref{sec:analytics} we develop an approximate analytical approach to
the barrier crossing for FBM, before drawing our conclusions in Sec.~\ref{sum}.
In the Appendices we describe the numerical algorithms used to generate
antipersistent and persistent FBM, and we validate in detail that these
truthfully produce FBM. We also briefly discuss the consistency of our results
for the case of a potential well, that is finite on both sides.

\section{Numerical analysis}
\label{sim}

In this Section we set up the Langevin description of FBM for external
Gaussian noise and present extensive simulations results for the barrier
crossing behavior.

\subsection{Langevin equation with fractional Gaussian noise}
\label{sec:strteqs}

We employ the overdamped Langevin equation for the position variable
$x(t)$ in the presence of an external potential $V(x)$,
\begin{equation}
\frac{dx(t)}{dt}=-\frac{1}{m\gamma}\frac{dV(x)}{dx}+\sqrt{D}\xi_H(t),
\label{LE}
\end{equation}
where $m$ is the particle mass, $\gamma$ the friction constant, $\xi_H(t)$ is
the fractional Gaussian noise, and $D$ is its intensity. The chosen initial
condition is $x(0)=0$. To study the activated escape from a potential well,
in what follows we use an harmonic potential of the form
\begin{equation}
\label{pot}
V(x)=\left\{\begin{array}{ll}
\frac{a}{2}x^2, & -\infty<x\le\sqrt{2}\\
-\infty, & x>\sqrt{2}\end{array}\right..
\end{equation}
with a truncation at positive $x=\sqrt{2}$, compare Fig.~\ref{potential}.
We note that we compared our simulations for the potential (\ref{pot}) to
the escape from an harmonic potential with symmetric truncation,
\begin{equation}
\label{pot2}
V(x)=\left\{\begin{array}{ll}
\infty, & -\infty<x<-\sqrt{2}\\
\frac{a}{2}x^2, & -\sqrt{2}\le x\le\sqrt{2}\\
-\infty, & x>\sqrt{2}\end{array}\right.,
\end{equation}
finding qualitative agreement with the results reported herein with
respect to the dependence of the distribution of escape times and the
dependence of the mean escape time on Hurst exponent and noise strength,
see the Appendix.

In continuous time the fractional Gaussian noise $\xi_H(t)$ is
understood as a derivative of the FBM \cite{mandelbrot, Qian}. This
is a stationary Gaussian process with an autocorrelation function
that in the long time limit decays as
\begin{equation}
\label{eqContACF}\left\langle\xi_H(0)\xi_H(t)\right\rangle\sim
2H(2H-1)t^{2H-2},
\end{equation}
for $0<H<1$, $H\ne 1/2$. Note that in the antipersistent
case, $0<H<1/2$, the autocorrelation function of the fractional
Gaussian noise is negative at long times. At $H = 1/2$ we have a
delta-correlated white noise. In a discrete time approximation used
in numerical simulations below the autocorrelation function of the
noise reads \cite{Qian}
\begin{equation}
\label{eqDiscreteACF} \left\langle \xi_H(0) \xi_H(n) \right\rangle =
\left[(n+1)^{2H}-2n^{2H}+(n-1)^{2H}\right].
\end{equation}
The continuum approximation (\ref{eqContACF}) is obtained from
Eq.~(\ref{eqDiscreteACF}) in the limit of large $n$ and identifying $n
\rightarrow t$. In what follows in analytical calculations and
numerical simulations we use the PDF of the fractional Gaussian noise
\begin{equation}
\phi(\xi_H)=\frac{1}{\sqrt{4\pi}}\exp\left(-\frac{\xi_H^2}{4}\right)
\end{equation}
with variance 2.

Replacing $x\rightarrow\left(m\gamma/a\right)^{H+1}x$ and $t\rightarrow
\left(m\gamma/a\right)t$ we pass to reduced variables:
\begin{equation}
\frac{dx(t)}{dt}=-x+\sqrt{D}\xi_H(t). \label{LEdl}
\end{equation}
The time-discretized version of Eq.~(\ref{LEdl}) acquires the form
\begin{equation}
x_{n+1} - x_n = -x_n \delta t + \sqrt{D} \delta t\xi_H(n),
\label{LE_discr}
\end{equation}
where $\delta t$ is a finite time step.

We applied the methods described in Refs.~\cite{fGnAlg1} and
\cite{fGnAlg2} for simulating fractional Gaussian noise with $H<1/2$
and $H>1/2$, respectively, as detailed in Appendix~\ref{appendixA}.
In the simulations the Hurst index $H$ was varied within the range
$[0.1, 0.85]$, whereas the noise intensity $D$ covered values from
1/6 to 1/2. Correspondingly, the escape time was varying in a range
covering three orders of magnitude.

\subsection{Numerical results for FBM Kramers escape}
\label{sec:numres}

\begin{figure}
\includegraphics[width=8.8cm]{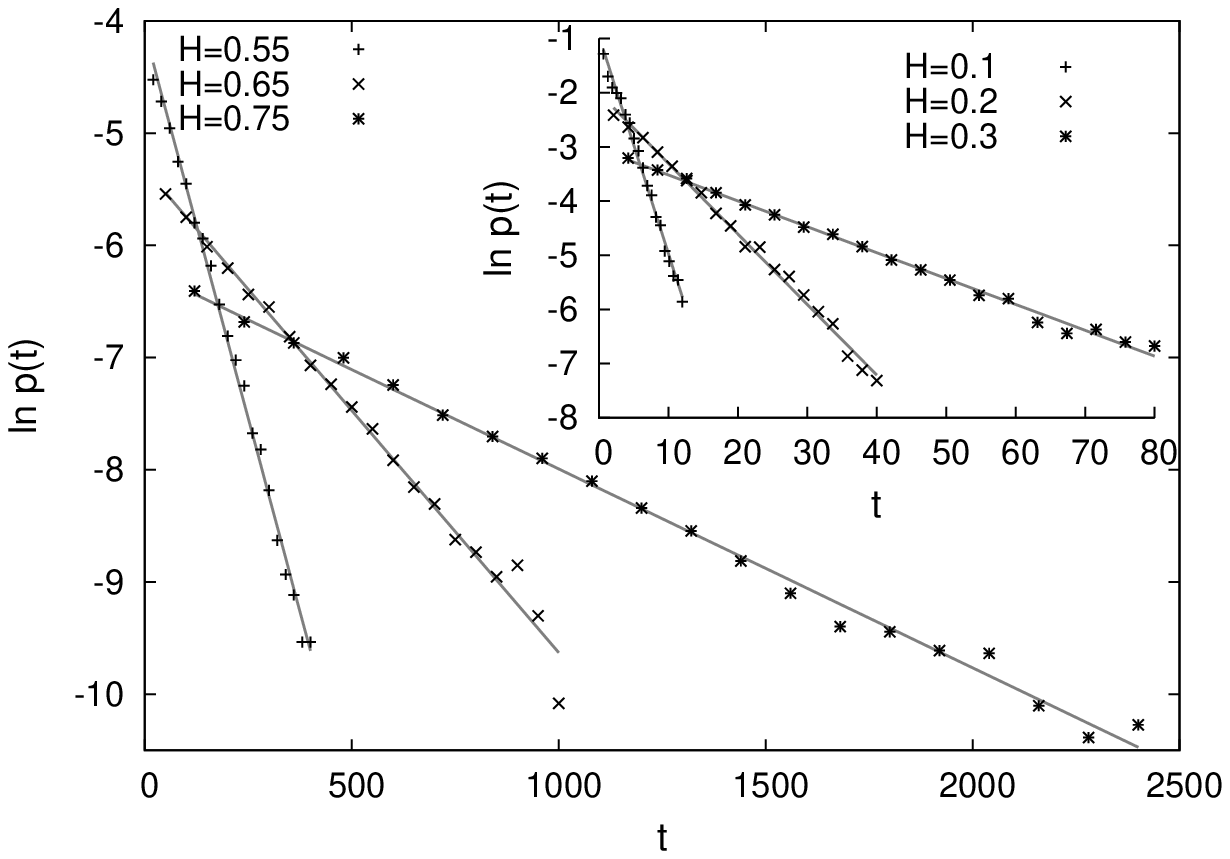}
\includegraphics[width=8.8cm]{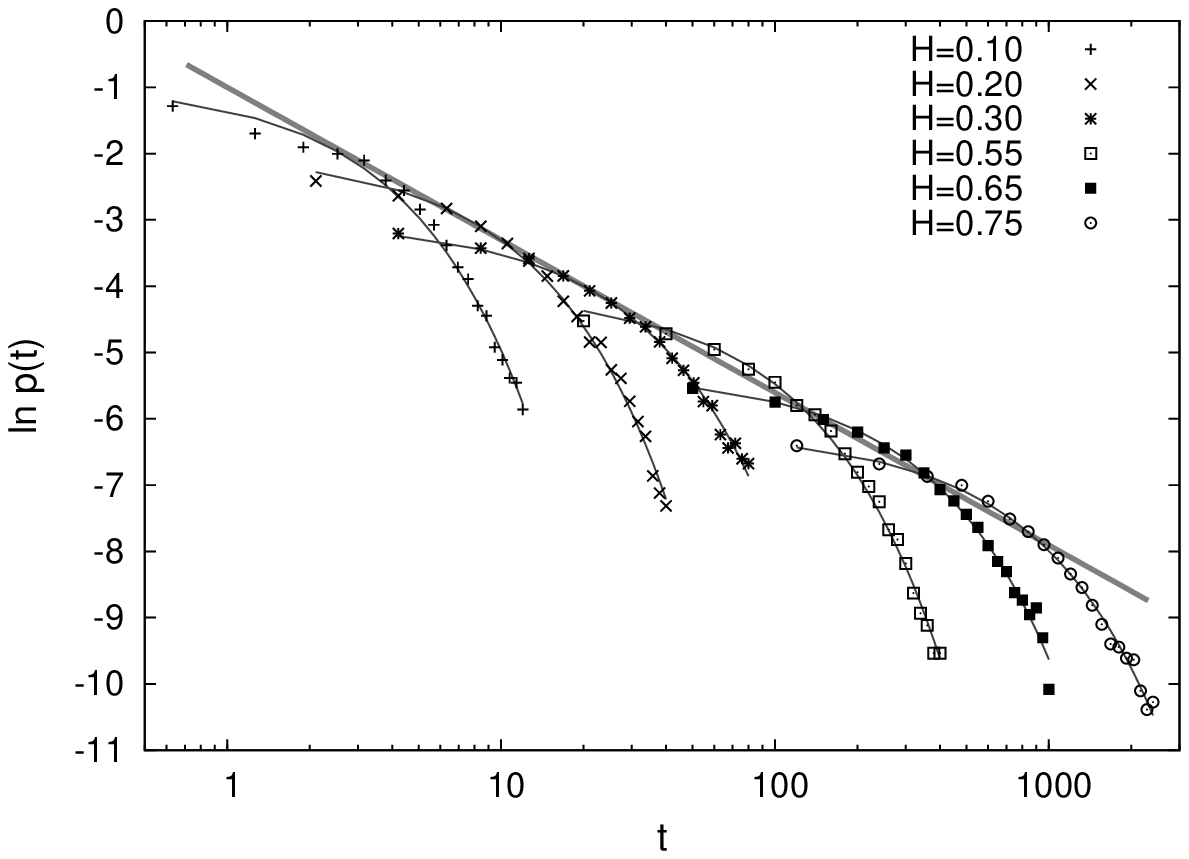}
\caption{\label{fig2}
Top panel: Probability density function (PDF) of the escape time nicely
demonstrating the exponential decay given by Eq.~(\ref{decay}). The main
plot depicts the persistent case ($H=0.55$, 0.65, 0.75), while in the
inset we show the anti-persistent case ($H=0.1$, 0.2, and 0.3). Here we
used the following simulations parameters: in the antipersistent case the
time increment is $\delta t=0.001$, the number of samples $N_\mathrm{stat}=
10^5$, the number of data points per sample $N_\mathrm{max} $ varied from
$2^{13}\approx8.2\times10^3$ to $2^{21}\approx2.1\times10^6$, and finally
the noise strength $D=0.25$; in the
persistent case we used $\delta t=0.001$, $N_\mathrm{stat}=20000$, $N_
\mathrm{max}=2^{20}\approx 10^6$, and $D=0.25$. Bottom panel: PDF of the
escape time
in a log-log representation. The decay curves have a common envelope $1/
(et)$ depicted by the straight grey line, see text.}
\end{figure}

In our simulations we follow the motion of the test particle governed by
the discrete Langevin equation (\ref{LE_discr}) in the harmonic potential
with one-sided truncation, Eq.~(\ref{pot}). Once the particle crosses the
point $x=\sqrt{2}$ it is removed, and the next particle started. This setup
is depicted in Fig.~\ref{potential}.

We first focus on the probability density function (PDF) of the first
escape time from the potential well.
In Fig.~\ref{fig2} we demonstrate that, in analogy to the classical case ($H=
1/2$) the probability density function (PDF) of the first escape time decays
exponentially with time, see Eq.~(\ref{decay}). This exponential decay is
observed nicely in the simulations data over the entire range of the Hurst
exponent. In the double-logarithmic plot in the bottom panel of Fig.~\ref{fig2}
one can see a common envelope of the curves for all values of $H$. Indeed the
shoulders of the individual exponential PDFs are located at points in time
where $t=T$, i.e., where the value of the PDFs is exactly $1/(eT)$. This is
the straight line plotted in Fig.~\ref{fig2}, showing good agreement, with a
slight underestimation for persistent Hurst exponents.

\begin{figure}
\includegraphics[width=8.8cm]{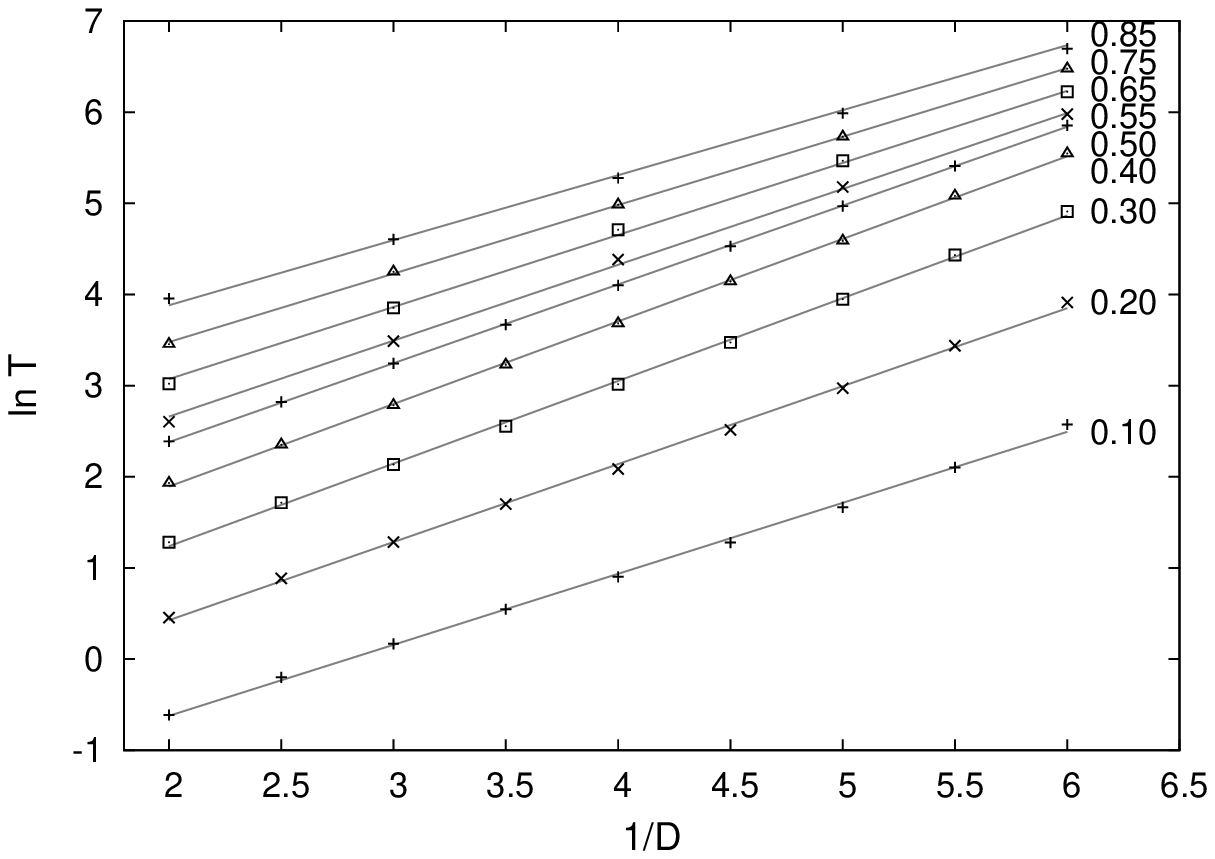}
\includegraphics[width=8.8cm]{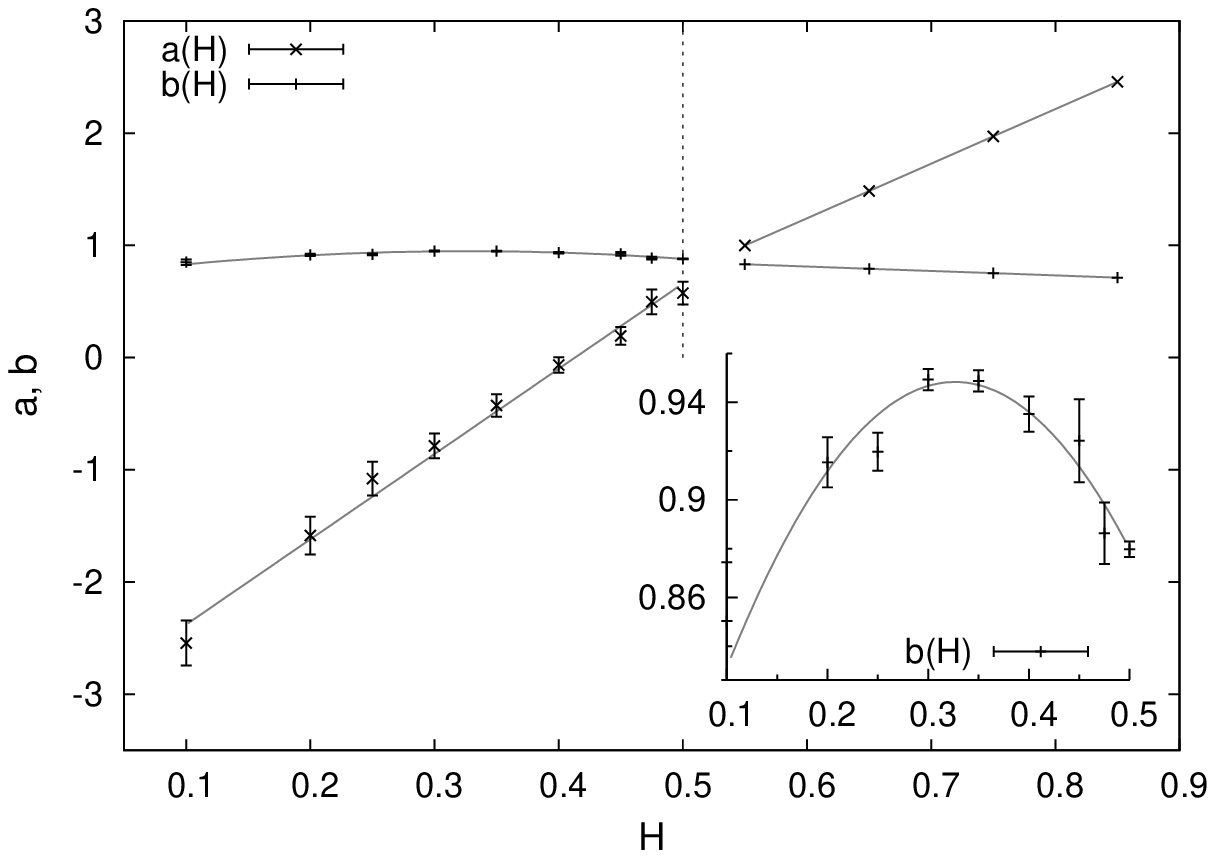}
\caption{\label{fig3} Top panel: Mean escape time $T$ as function of
inverse noise intensity $1/D$ in logarithmic scale. The simulations
results are shown for Hurst exponents $H=0.85$, 0.75, 0.65, 0.55,
0.5, 0.4, 0.3, 0.2, and 0.1 (top to bottom). The solid lines
represent a linear fit. Bottom panel: Fitting coefficients $a(H)$
and $b(H)$ from Eq.~(\ref{Tfitting}). The inset shows $b(H)$ for the
antipersistent case at higher resolution. The symbols represent the
values of $a$ and $b$ from the simulations, while the solid lines
show the respective fits with Eqs.~(\ref{a1fit}) to (\ref{b2fit}).
Simulations parameters: in the antipersistent case we used the time
increment $\delta t= 0.001$, number of samples
$N_\mathrm{stat}=10^5$, and the data points per sample
$N_\mathrm{max} $ varied from $2^{13}\approx8.2\times10^3$ to
$2^{21}\approx2.1\times10^6$; in the persistent case $\delta t$
ranges from 0.01 to 0.001, $N_\mathrm{stat}=10^6$, and
$N_{\mathrm{max}}$ varied from $2^{13} \approx8.2\times10^3$ to
$2^{18}\approx2.6\times10^5$.}
\end{figure}

\begin{figure}
\includegraphics[width=8.8cm]{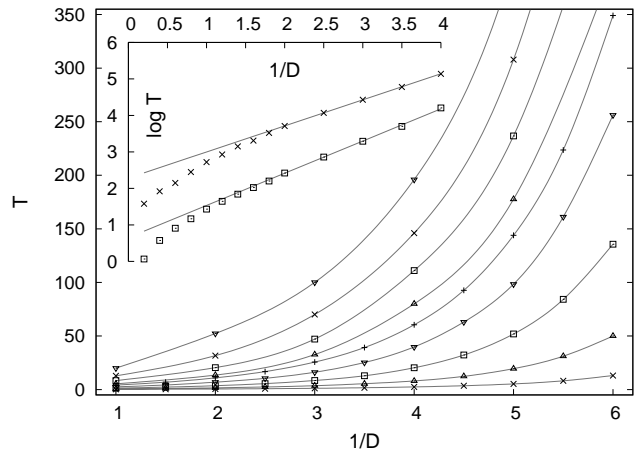}
\caption{Mean escape time $T$ as function of inverse noise intensity $1/D$ in
linear scale. Data points are the same as in Fig.~\ref{fig3}. Again, the dots
represent the simulations data, and the solid lines depict the exponential fit
with the global fitting coefficients according to Eqs.~(\ref{a1fit}) to
(\ref{b2fit}). From top to bottom: Hurst exponent $H=0.85$ ($\triangledown$),
$H=0.75$ ($\times$), $H=0.65$ ($\square$), $H=0.55$ ($\triangle$), $H=0.50$
($+$), $H=0.40$ ($\triangledown$), $H=0.30$ (($\square$), $H=0.20$ ($\triangle
$), $H=0.10$ ($\times$).
Inset: Too high noise intensity violating the high barrier
assumption leads to deviations from the exponential behavior. Hurst exponent
$H=0.5$ ($\square$) and $H=0.8$ ($\times$).}
\label{linear}
\end{figure}

In Fig.~\ref{fig3} we demonstrate that the mean escape time $T$ follows an
exponential behavior as function of the inverse noise intensity, $1/D$, in
analogy to the classical Kramers case. We observe that in both persistent
and antipersistent cases this functional dependence may be approximated by
a linear fit of the form
\begin{equation}
\label{Tfitting}
\ln T\left(D;H\right)=a(H)+\frac{b(H)}{D},
\end{equation}
where both fitting coefficients $a$ and $b$ are functions of the Hurst
exponent $H$. These, in turn, show different behavior for antipersitence
and persistence of the motion:

(i) In the persistent case $1/2\le H<1$ both coefficients are linear functions
of the Hurst exponent. We found empirically from best fits that
\begin{eqnarray}
\label{a1fit}a\left(H\geq1/2\right)&=&a_1+a_2H,\\
\label{b1fit}b\left(H\geq1/2\right)&=&b_1+b_2H,
\end{eqnarray}
where $a_1=- 1.680$, $a_2=4.869$, $b_1=1.051$, and $b_2=-0.399$. The good
quality of this linear description is seen in the bottom panel of
Fig.~\ref{fig3}.

(ii) Contrasting this behavior, in the antipersistent case the coefficient
$a\left(H\right)$ is still well described by a linear $H$-dependence, while
$b\left(H\right)$ is well represented by a parabolic dependence:
\begin{eqnarray}
\label{a2fit}a\left(H\leq1/2\right)&=&\tilde{a}_1+\tilde{a}_2H,\\
\label{b2fit}b\left(H\leq1/2\right)&=&\tilde{b}_1+\tilde{b}_2H+\tilde{b}_3H^2.
\end{eqnarray}
The best fit parameters are determined as $\tilde{a}_1=-3.019$, $\tilde{a}_2=
7.296$, $\tilde{b}_1=0.705$, $\tilde{b}_2=1.490$ and $\tilde{b}_3=-2.281$.
Again, Fig.~\ref{fig3} demonstrates good agreement with this chosen
$H$-dependence. In Fig.~\ref{linear} we show the quality of these fits
(solid curves) on a linear scale. Note the deviations from the exponential
behavior when the noise intensity becomes too large [in our simulation for
values $D>1$]. In that case the high barrier limit is violated and the
results obtained herein are no more applicable, in correspondence to regular
Brownian barrier crossing behavior.

The general agreement with the law (\ref{Tfitting}) is excellent, keeping in
mind that the error of the simulations data is of the magnitude of the points.
Remarkably the characteristic escape time increases from low to high Hurst
exponent. In other words, the less persistent motion shows the faster
escape. This observation is consistent throughout our simulations. In
particular this behavior is not qualitatively changed for a parabolic
potential of the type (\ref{pot2}) with symmetric cutoff.

\begin{figure}
\includegraphics[width=8.8cm]{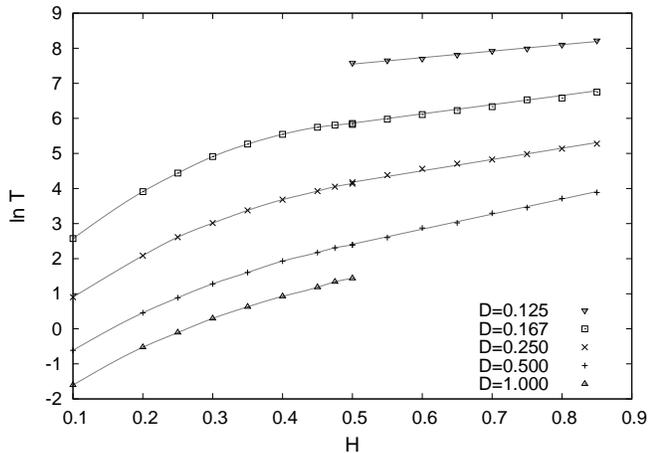}
\caption{\label{fig4} Mean escape time as a function of the Hurst exponent
for different noise intensities $D$. The solid lines correspond to the fits
used in Fig.~\ref{fig3}, converted according to Eqs.~(\ref{lnT(H)anti}) and
(\ref{lnT(H)pers}).}
\end{figure}

In Fig.~\ref{fig4} the mean escape time is reanalyzed as a function of the
Hurst exponent. In accordance with the results presented in Fig.~\ref{fig3},
there is a parabolic dependence of $\ln T$ versus $H$ in the antipersistent
case ($0<H<1/2$),
\begin{equation}
\label{lnT(H)anti}
\ln T=\tilde{c}_1+\tilde{c}_2H+\tilde{c}_3H^2,
\end{equation}
where $\tilde{c}_1=\tilde{a}_1+\tilde{b}_1/D$, $\tilde{c}_2=\tilde{a}_2+
\tilde{b}_2/D$, and $\tilde{c}_3=\tilde{b}_3/D$. In the persistent case $1/2
<H<1$ the relation is linear, corresponding to
\begin{equation}
\label{lnT(H)pers}
\ln T={c}_1+{c}_2H,
\end{equation}
where $c_1=a_1+b_1/D$ and $c_2=a_2+b_2/D$. The agreement with the
fit function is favorable, and the continuation between
antipersistent and persistent cases appears relatively smooth. The
latter supports the good convergence of the simulations algorithms
used in the antipersistent and persistent regimes (see Appendix A).
At the same time the difference between the behaviors in the
two regimes (persistent versus antipersistent) is quite
distinct.

\begin{figure}
\includegraphics[width=8.8cm]{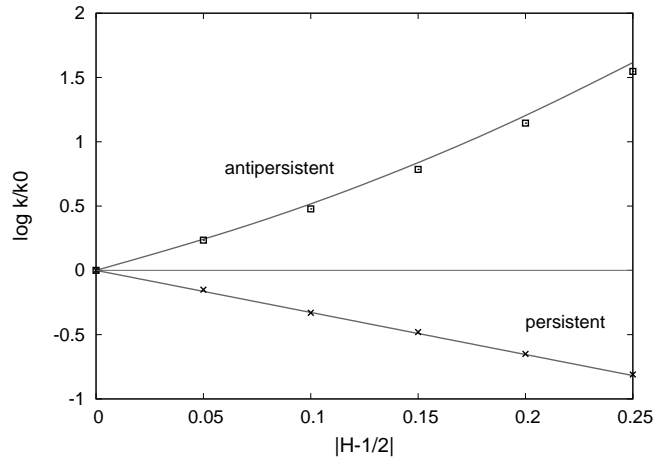}
\caption{\label{fig5} Relative escape rate $k\left(H\right)/k\left(H=1/2
\right)$ as a function of $\left|H-1/2 \right|$ for the persistent and
antipersistent cases for $D=0.25$. The solid parabolic and straight lines are obtained
from fits to the numerical data according to Eqs.~(\ref{lnT(H)anti}) and
(\ref{lnT(H)pers}), respectively. A fine coincidence is observed. }
\end{figure}

Fig.~\ref{fig5} shows an alternative way to represent the behavior from
Figs.~\ref{fig3} and \ref{fig4}, namely, in terms of the ratio $k\left(H
\right)/k\left(H=1/2\right)$ of the escape rates (that is, the inverse mean
escape times) as function of the deviation $\left|H-1/2\right|$ from
normal diffusion at $H=1/2$.
The rates increase with decreasing Hurst exponent, i.e., the less persistent
the motion is the higher becomes the corresponding rate. One can also see the
difference between the parabolic dependence in the antipersistent case and
the linear relation for persistent motion.

\begin{figure}
\includegraphics[width=8.8cm]{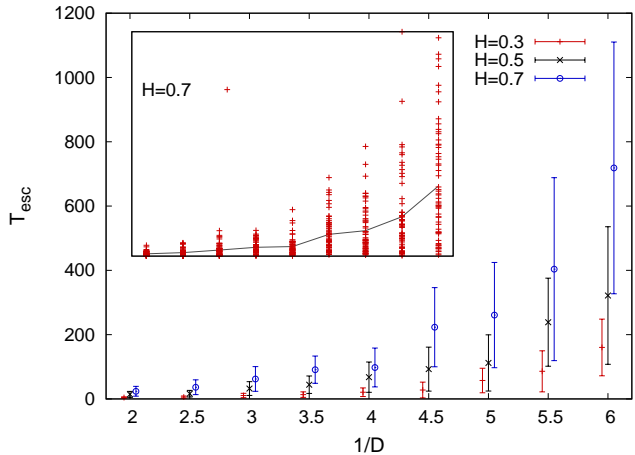}
\caption{(Color online) Results for the mean escape time. Main
graph: for three different values of the Hurst exponents ($H=0.3$,
0.5, and 0.7) we show the average value and the standard deviation
around that value for different inverse noise intensities $1/D$.
Note that $T_{esc}$ is evaluated at $1/D=2$, 2.5, $3,\ldots$.
at each $H$ value. In the figure the results for different $H$ at each
given value of $1/D$ are slightly shifted with respect to each other for
illustrative purposes.
Inset: values of the escape times for each
individual out of 60 trajectories for $H=0.7$. The grey line
connects the averages. In the simulations we used the time increment
$\delta t=0.001$ and the number of points per trajectory
$N_\mathrm{max}$ varied from $2^{11} \approx2\times10^3$ to
$2^{20}\approx10^6$.} \label{scatter}
\end{figure}

Finally in Fig.~\ref{scatter} we explore the distribution of the results for
the mean escape time between different samples of only 60 trajectories. Again
we see the increased escape time at higher Hurst exponent. We also clearly
observe that the variation around the average values increases significantly
for higher Hurst exponent. In particular the noise for the plotted case
$H=0.3$ is consistently smaller than for the Brownian limit $H=1/2$.

\section{Analytical approach to FBM driven Kramers escape}
\label{sec:analytics}

In this Section we derive analytical results for the escape behavior driven
by fractional Gaussian noise. In particular we concentrate on the mean escape
time and the autocorrelation function for FBM in an harmonic potential. We
compare the results to the numerical findings from the preceding Section.

\subsection{Wilemski-Fixmann approximation}
\label{sec:W-F approx}

The investigation of first passage times for non-Markovian processes has
a long history in mathematical literature, for instance, see Refs.~\cite{%
Slepian,Molchan,Rice,stratonovich}, and appears in different fields of
science, including chemical physics \cite{haenggi}, polymer physics
\cite{willemski,szabo} and neuroscience \cite{Igor-Tatiana}. However, no
general theory exists for such processes, and different approximations are
used depending on whether the process is Gaussian or not, whether its
trajectories are differentiable or not, \emph{etc}. Our analytical approach
to the escape problem considered herein is based on a special case of the
Wilemski-Fixmann approximation (WFA)\cite{willemski} used in polymer physics
\cite {szabo}. As shown in Ref.~\cite{Igor_PRL} the application of the WFA to
a first passage problem corresponds to a renewal approximation
\cite{redner,hughes} in which, however, the correct Green's functions of the
original non-renewal processes are used. The WFA is essentially a first
approximation in the perturbative series derived by Likthman and Marques
\cite{Marques}, while higher approximations lead to quite involved
expressions.

Our theoretical approach starts from the relation
\begin{eqnarray}
\label{eq_1_1_} \nonumber
G(x,t|x_0,0)&=&\delta(x-x_0)\delta(t)\\
&&\hspace*{-1.6cm}+\int_0^tF(x,t',x_0,0)G(x,t|x,t')dt',
\end{eqnarray}
where $G(x,t|x_0,0)$ is the conditional probability to find the particle at
position $x$ at time $t$, provided that it started at $x_0$ at time $t=0$.
Moreover $F(x,t,x_0, 0)$ represents the first passage time PDF to cross the
distance $|x-x_0|$ during the time interval $t$, and $G\left(
x,t|x,t'\right)$ is the conditional probability to be at $x$ at time $t$,
provided $x$ was visited earlier at time $t'$. If the inequality $x_0\ne x$
holds the $\delta$-term can be omitted. For a continuous Markovian process
Eq.~\eqref{eq_1_1_} is exact. Its meaning
is that a particle, having started at $x_0$ at time 0 and being at a site
$x$ at time $t$, might have visited $x$ at some time $t'$ before, departed
from $x$, and returned \cite{redner,hughes}. For the non-Markovian case
Eq.~\eqref{eq_1_1_} neglects the correlations in the motion of the particle
before and after the first passage through the point $x$. Such correlations
lead to the dependence of the return probability (expressed through $G(x,t|
x,t')$) on the pre-history \cite{Igor_PRL}, and can be taken into account
systematically in higher order approximations involving multi-point
distribution functions \cite{Marques}. The approximation given by
Eq.~\eqref{eq_1_1_} may become incorrect in the case of strongly correlated
(persistent) processes. In that case our numerical results still show
exponential first passage time behavior corresponding to a finite
mean first passage time, while the WFA breaks down, as will be shown
below.

To proceed recall that according to Bayes' formula, $G(x,t|x_0,0)=P(x,t;x_0,
0)/P(x_0,0)$ and $G(x,t|x,t')=P(x,t;x,t')/P(x,t')$. Here $P(x,t;x,0)$ and
$P(x,t)$ are the corresponding two- and one-point probability densities.
Eq.~\eqref{eq_1_1_} can therefore be rewritten in the form
\begin{eqnarray}
\label{eq_1_2_} \nonumber
P(x,t;x_0,0)&=&P(x_0,0)\\
&&\hspace*{-1.6cm}\times\int_0^tF(x,t',x_0,0)\frac{P(x,t;x,t')}{P(x,t')}dt'.
\end{eqnarray}
Integration with respect to $x_0$ in Eq.~\eqref{eq_1_2_} leads to
the expression
\begin{equation}
\label{eq_1_3_}
P(x,t)=\int_0^tF(x,t')\frac{P(x,t;x,t')}{P(x,t')}dt',
\end{equation}
where
\begin{equation}
\label{eq_1_4_}
F(x,t')=\int_{-\infty}^{\infty}P(x_0,0)F(x,t',x_0,t)dx_0.
\end{equation}
Thus, the first escape PDF $F$ is obtained as an average over the
initial distribution.

In what follows we make use of the fact that in our numerical simulations
the typical relaxation times for a particle in an harmonic potential well
are much shorter than the typical mean escape times. Therefore the random
process $x(t)$ can be considered as stationary, that is, $P(x,t)=P_{st}(x)$
and $P(x,t;x,t')=P(x,x,t-t')$. Transferring $P$ from the left hand side to
the right of Eq.~\eqref{eq_1_3_} we find
\begin{equation}
\label{eq_1_5_} 1=\int_0^tF(x,t')\frac{P(x,x,t-t')}{P_{st}^2(x)}dt'.
\end{equation}
This relation converts to an algebraic equation after Laplace transformation,
\begin{equation}
\label{eq_1_6_}
\frac{1}{s}=\tilde{F}(x,s)\frac{\tilde{P}(x,x,s)}{P_{st}^2(x)}.
\end{equation}
Here we express the Laplace transform of a function $f(t)$ as $\tilde{f}(s)=
\int_0^{\infty}f(t)\exp(-st)dt$. Since $P(x,x,t\to\infty)\to P_{st}^2(x)$, we
see that $\tilde{P}(x,x,s\to 0)\to P_{st}^{2}(x)/s$, and for small $s$ we may
expand $\tilde{P}(x,x,s)$ in the form
\begin{equation}
\label{eq_1_7_}
\tilde{P}(x,x,s)\approx\frac{P_{st}^2(x)}{s}+A(x)+O(s),
\end{equation}
where we use the abbreviation
\begin{eqnarray}
\nonumber
A(x)&=&\lim_{s\to0}\left[\tilde{P}(x,x,s)-\frac{P_{st}^2(x)}{s}\right]\\
&=&\int_0^{\infty}\Big[P(x,x,t)-P_{st}^2(x)\Big]dt.
\label{eq_1_8_}
\end{eqnarray}
After inserting Eq.~\eqref{eq_1_7_} into Eq.~\eqref{eq_1_6_} we get
\begin{eqnarray}
\nonumber
\tilde{F}(x,s)=\frac{P_{st}^2(x)}{s\tilde{P}(x,x,s)}&\approx&
\frac{P_{st}^2(x)}{P_{st}^2(x)+A(x)s}\\
&\approx&1-\frac{A(x)}{P_{st}^2(x)}s+\ldots.
\label{eq_1_9_}
\end{eqnarray}
Thus, with the use of Eq.~\eqref{eq_1_8_}, we find
\begin{eqnarray}
\nonumber
T(x)&=&-\left.\frac{d}{ds}\tilde{F}(x,s)\right|_{s=0}\\
&=&\frac{A(x)}{P_{st}^2(x)}=\int_0^{\infty}\left[\frac{P(x,x,t)}{P_{st}^2(x)}
-1\right]dt.
\label{eq_1_10_}
\end{eqnarray}
We will use this result below.

Before proceeding two remarks are in order: First, we note that in the theory
developed here we use the ensemble average over initial values $x_0$, while
in the simulations we use $x_0=0$ for all trajectories. Nevertheless, we can
employ Eq.~\eqref{eq_1_9_} since typically the relaxation time is much shorter
than the mean escape time and, therefore, the system quickly converges to the
stationary state, which is independent of the initial condition. And second,
when writing Eq.~\eqref{eq_1_7_} we implicitly assume that the mean escape
time exists. This is in accordance with the numerical observation that the
escape time PDF has the simple exponential form (\ref{decay}).

\subsection{Mean escape time for Gaussian processes}
\label{sec:METGauss}

To proceed we exploit the Gaussian property of FBM processes. We recall
the expressions for one- and two-point Gaussian PDFs, namely,
\begin{equation}
\label{eq_2_1_}
P_{st}(x)=\frac{1}{\sqrt{2\pi\sigma^2}}\exp\left(-\frac{x^2}{2\sigma^2}\right),
\end{equation}
where $\sigma^2=\left\langle x^2\right\rangle_{st}$ is the variance in the
stationary state of a particle in an harmonic potential well. Moreover
\begin{eqnarray}
\nonumber
P(x,y,t)&=&\frac{1}{2\pi\sigma_x\sigma_y\sqrt{1-g^2(t)}}\\
&&\hspace*{-1.4cm}\times\exp\left\{-\frac{1}{2(1-g^2)}\left(\frac{x^2}{\sigma
_x^2}+\frac{y^2}{\sigma_y^2}-\frac{2gxy}{\sigma_x\sigma_y}\right)\right\},
\label{eq_2_2_}
\end{eqnarray}
where $g(t)$ is the normalized autocorrelation function in the stationary
state,
\begin{equation}
\label{eq_2_3_}
g(x,y,\tau)=\frac{\left\langle x(t)y(t+\tau)\right\rangle_{st}}{\sigma_x
\sigma_y}  .
\end{equation}
Thus, within our approximation
\begin{equation}
\label{eq_2_4_}
P(x,x,t)=\frac{1}{2\pi\sigma^2\sqrt{1-g^2}}\exp\left\{-\frac{x^2}{\sigma^2
\left(1+g\right)}\right\},
\end{equation}
and we obtain the mean time
\begin{equation}
T=\int_0^{\infty}\left\{\frac{1}{\sqrt{1-g^2(\tau)}}\exp\left[\frac{x^2}{\left
\langle
x^2\right\rangle_{st}}\frac{g(\tau)}{1+g(\tau)}\right]-1\right\}d\tau.
\label{eq_2_5_}
\end{equation}
Here we identified
\begin{equation}
\label{eq_2_6_}
g(\tau)=\frac{\left\langle x(t)x(t+\tau)\right\rangle_{st}}{\left\langle
x^2\right\rangle_{st}}.
\end{equation}
Expressions $\langle x(t)x(t+\tau)\rangle_{st}$ and $\langle x^2\rangle_{st}$
are calculated in App.~\ref{appendixC}.

\subsection{Persistent and antipersistent cases}
\label{sec:divergence}

Consider now the asymptotic behavior of the integrand in expression
(\ref{eq_2_5_}) at $\tau\rightarrow\infty$,
\begin{eqnarray}
\nonumber
\left\{...\right\}&\mathop{\approx}\limits_{\tau\to\infty}&\left[1+g^2(\tau)
\right]\left[1+\frac{x_{esc}^{2}}{\left\langle
x^2\right\rangle_{st}}\frac{
g(\tau)}{1+g(\tau)}\right]-1\\
&\approx&\frac{x_{esc}^2}{\left\langle x^2\right\rangle_{st}}g(\tau).
\label{eeq_7_}
\end{eqnarray}
Since $g(\tau)\sim\tau^{2H-2}$, the integrand decays slowly; the integral
in Eq.~(\ref{eq_2_5_}) itself converges for $H<1/2$ and diverges for $H>1/2$.

\begin{figure}
\includegraphics[width=8.8cm]{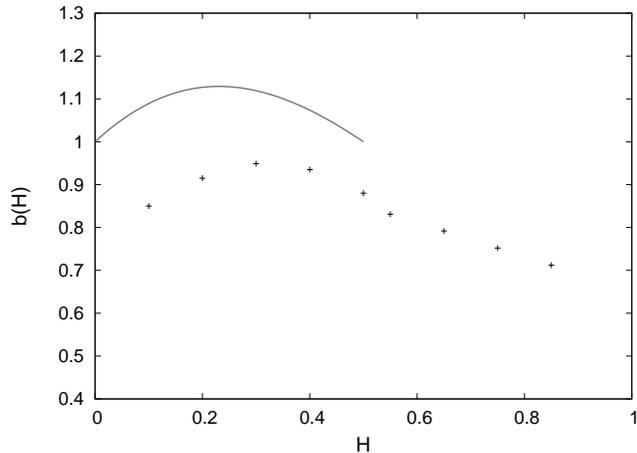}
\caption{\label{fig51} Coefficient $b(H)$ occurring in the empirical formula
(\ref{Tfitting}) for the mean escape time. Symbols: Values obtained from best
fit. Solid line: Theoretical behavior described by Eq.~\eqref{EQ__4}.}
\end{figure}

\begin{figure}
\includegraphics[width=8.8cm]{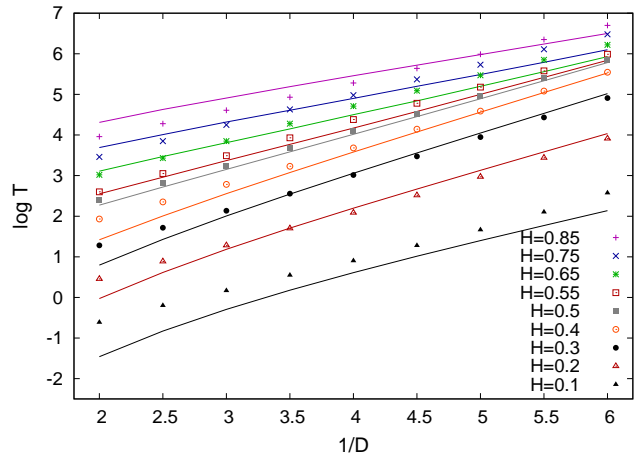}
\caption{\label{fig6} (Color online.)
Mean escape time $T$ as function of inverse noise
intensity, $1/D$. Symbols: simulations results. The solid lines show the
analytical result given by Eqs.~(\ref{eq_2_5_}) and (\ref{eq_2_6_})
in the antipersistent case $0<H<1/2$. For persistent motion $1/2<
H<1$ the solid lines represent a fit by Eqs.~(\ref{eq_2_5_}) and
(\ref{eq_2_6_}) based on numerical
truncation of the integral in Eq.~(\ref{eq_2_5_}) with cutoff time
$\tau_{cut}=18$.}
\end{figure}

Focusing at first on the antipersistent case we notice that according to
Eq.~\eqref{eq_2_5_} the main contribution comes from the integrand estimated
at $g(\tau)\sim1$, which immediately leads to
\begin{equation}
\label{EQ__3} T\simeq\exp\left(\frac{1}{\left\langle
x^2\right\rangle_{st}}\right),
\end{equation}
being a kind of generalization of the standard transition-state
arguments to the FBM case. Recalling that for our harmonic
potential, $\left\langle x^2\right\rangle_{st} =D\Gamma(2H+1)$, we
obtain an estimate for the coefficient $b(H)$ in the empirical
formula for the escape time, Eq.~(\ref{Tfitting}). Namely, we find
\begin{equation}
\label{EQ__4} b(H)=\frac{1}{\Gamma(2H+1)}.
\end{equation}
Eq.~\eqref{EQ__4} provides a surprisingly good approximation to the behavior
of $b(H)$ obtained from the simulations, as shown in Fig.~\ref{fig51}. In
particular, approximation \eqref{EQ__4} shows the nontrivial maximum for
intermediate $H$-values. Fig.~\ref{fig6} shows the values for the mean escape
time obtained from our simulations of the antipersistent process with $0<H<1
/2$, along with the behavior predicted by Eqs.~(\ref{eq_2_5_}) and
(\ref{eq_2_6_}).

In the persistent case the integral in expression (\ref{eq_2_5_}) diverges.
We show that a suitable truncation at some upper bound $\tau_{cut}$ leads to a
quite good agreement with the behavior recovered from simulations. Physically
such a truncation always exists due to the finiteness of the slow power-law
decay of the autocorrelation function of fractional Gaussian noise. Thus, we
would always expect finite mean escape times also in the persistent range.
Because of the slow divergence of the integral for $T$, we may expect a weak
dependence of the integral on the cutoff parameter $\tau_{cut}$ if only it is
chosen large enough. Indeed, we found in our numerical simulations that the
value $\tau_{cut}=18$ already gives good agreement with the numerical
simulation, see Fig.~\ref{fig6} bottom.

\section{Summary}
\label{sum}

In this work we present an extensive analysis of the generalized Kramers
escape from a potential well for a particle subject to fractional Brownian
motion. Specifically we considered a particle whose motion is governed by
the Langevin equation driven by external fractional Gaussian noise. The
motion we consider is thus not subject to the fluctuation dissipation
theorem. Potential applications for such behavior may,
for instance, include geo- and astrophysical fluctuations, stock market
pricing, or teletraffic.

Based on simulations and analytical derivations we showed that, despite the
driving fractional Gaussian noise, the escape dynamics preserved the classical
exponential shape of the distribution of escape times. Deviations from the
behavior for regular Gaussian white noise
are found in the activation dependence of the mean escape
time on the noise intensity at different values of the Hurst exponent $H$.

The escape turns out to slow down for increasing value of the Hurst exponent.
Thus in the persistent case $1/2<H<1$ the escape is slower than in the
antipersistent case $0<H<1/2$, and the latter is faster than for ordinary
Brownian case. This somewhat surprising result is in accordance with previous
results for the first passage time \cite{Molchan}, where the scaling
exponent of the first passage time distribution decreases for increasing $H$.
We note that this observation is not restricted to the asymmetrically truncated
harmonic potential used in this work, but also occurs for a symmetric
truncation of the harmonic potential at $x=\pm\sqrt{2}$.

Analyzing the detailed behavior of the mean escape time we find that the
logarithm, $\log T$ in the entire simulations range $H=0.1,\ldots,0.85$
depends linearly on the inverse noise intensity, $1/D$. This activation
dependence is thus preserved for both antipersistent and persistent cases.
Conversely, the behavior of $\log T$ on the Hurst exponent shows a linear
dependence in the persistent case, while in the antipersistent case we
find a nonlinear dependence.

We note that fractional Brownian motion is an ergodic process in the sense
that time and ensemble averages coincide, albeit the convergence to ergodicity
is algebraically slow with the measurement time
\cite{deng}. For sufficiently long averaging
times the dynamic behavior of time and ensemble averages of
individual trajectories should therefore be identical. This
contrasts the behavior for continuous time random walk processes
with diverging characteristic waiting times \cite{web} or with
correlations in waiting times or jump lengths \cite{vincent1}.

The understanding of fractional Brownian motion in several aspects remains
formidable. We expect that this work contributes toward the demystification
of this seemingly simple stochastic process.

\begin{acknowledgments}
Discussions with Olivier Benichou, Jae-Hyung Jeon, Yossi Klafter,
Michael Lomholt, Vincent Tejedor, and Raphael Voituriez are
gratefully acknowledged. We also acknowledge funding from the
Deutsche Forschungsgemeinschaft within SFB 555 Research
Collaboration Program and the European Commission through a MC IIF
Grant No.219966 LeFrac.
\end{acknowledgments}

\begin{appendix}

\section{Description of FBM generators}
\label{appendixA}

Here we briefly describe the generators with which we simulated FBM. It
should be noted that the generators provide best results for either the
antipersistent case $0<H<1/2$ or for the persistent case $1/2<H<1$.

A fast and precise (see the tests in Appendix B) generator for
fractional Gaussian noise in the \textit{anti-persistent} case is
described in Ref.~\cite{fGnAlg1}. In brief, the idea is as follows.

First, we define a function
\begin{equation}
\label{Rx}
R_x(n)=\left\{\begin{array}{ll} 2^{-1} \left[1-( n/N_\mathrm{max})^{2H}
\right], & 0\leq n\leq N_\mathrm{max}\\[0.2cm]
R_x(2N_\mathrm{max}-n), & N_\mathrm{max}<n<2N_\mathrm{max}
\end{array}
\right.
\end{equation}
where $H$ is the Hurst parameter ($0<H<1/2$), $n$ is the number of steps
corresponding to time in the continuous time limit, and $N_\mathrm{max}$
is the length of the random sample. Second, we perform a discrete Fourier
transformation of Eq.~(\ref{Rx}), with $S_x(k)=F\{R_x(n)\}.$

We then define
\begin{equation}
\label{Xk}
X(k)=\left\{ \begin{array}{ll}
0, & k=0\\[0.2cm]
\exp(i\theta_k)\xi(k)\sqrt{S_x(k)}, & 0<k<N_\mathrm{max}\\[0.2cm]
\xi(k)\sqrt{S_x(k)}, & k=N_\mathrm{max}\\[0.2cm]
X^*(2N_\mathrm{max}-k), & N_\mathrm{max}<k<2N_\mathrm{max},
\end{array}
\right.
\end{equation}
where the symbol $*$ stands for complex conjugation, $\theta_k$ are uniform
random numbers from $[0,2\pi)$, and $\xi(k)$ are Gaussian random variables
with zero mean and variance equal to 2. All random variables are independent
of each other.

Finally, we set $y(n)=x(n)-x(0),$ where
$x(n)=F^{-1}\left\{X(k)\right\}$ is the inverse Fourier
transformation of Eq.~(\ref{Xk}). The quantity $y(n)$ represents a
free [i.e., in absence of an external force] fractional Brownian
trajectory which is to be differentiated with respect to time, to obtain
fractional Gaussian random numbers. Since the variance
$\left\langle\xi^2\right\rangle$ depends on the number of steps
$N_\mathrm{max}$, it is normalized such that $\left
\langle\xi^2\right\rangle=2$.

Despite the availability of several exact simulation methods, for the
persistent case we chose an approximate but efficient simulation method.
This generator exploits the spectral properties of fractional Gaussian
noise \cite{fGnAlg2}. The method uses the following steps:

(i) Take white Gaussian noise $\xi(t)$, where $t$ is an integer.

(ii) Calculate the spectral density of this Gaussian noise and perform a
Fourier transformation, $S(k)=F\{\xi(t)\}$.

(iii) Introduce correlations multiplying it by $1/k^{H-1/2}$, where
$1/2<H<1$.

(iv) Inverse Fourier transform
$\xi_H(t)=F^{-1}\{S(k)k^{1/2-H}\}$, to obtain approximate
fractional Gaussian noise with the index $H$.

(v) Normalize the noise.

In Appendix B we demonstrate that this method reliably produces FBM.

We note that since we approximate the integral representation, this creates
two types of errors, a `low frequency' one due to the truncation of the
limit of integration and a `high frequency' one caused by replacing the
integral by a sum. By using various tests, we estimated the best
discretization parameters. We used the maximum sample length of $2^{24}
\approx1.7\times10^7$
steps, the time increment varying within the interval $[0.001, 0.01]$.

\section{Testing the numerical algorithm}
\label{appendixB}

To check our simulations algorithm based on numerical integration of
the Langevin equation (\ref{LE_discr}) we performed a number of
tests to validate the FBM we create with the generators sketched in
Appendix A.

\begin{figure}
\includegraphics[width=8.8cm]{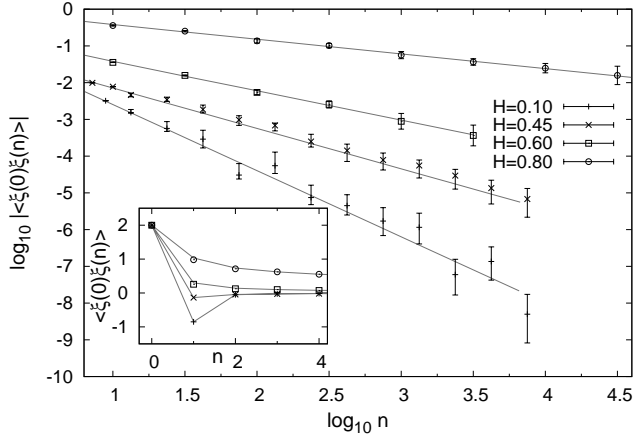}
\caption{\label{ACFfGn} Absolute value of the autocorrelation function of
fractional Gaussian noise for the entire range of the Hurst exponent $H$ in a
log-log
scale as function of the number of time steps $n$. Inset: autocorrelation
functions with the same Hurst indices for small numbers of steps
$n$ on a linear scale. The numerical results are shown
for $H=0.10$, 0.45, 0.60, and 0.80.
The solid lines in the main graph correspond to the analytical solution
(\ref{eqDiscreteACF}). Simulations parameters: number of simulated
samples $N_\mathrm{stat}=20,000$, each of length $N_\mathrm{max}=2^{13}
\approx8.2\times10^3$ for the antipersistent case, and $2^{15}\approx3.3\times
10^4$ for the persistent case, respectively.}
\end{figure}

First, we calculated the autocorrelation function of the
fractional Gaussian noise. As shown in Fig.~\ref{ACFfGn}, the simulated
data show excellent agreement with the analyical result (solid lines) given
by Eq.~(\ref{eqDiscreteACF}) for discrete time steps.

\begin{figure}
\includegraphics[width=8.8cm]{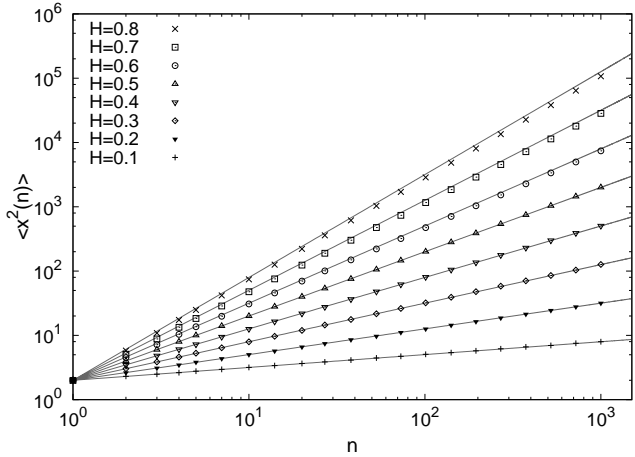}
\caption{\label{msd1} Mean squared displacement for free FBM in log-log
representation. The solid lines show the analytical expression
(\ref{msd-discr}) while the symbols depict the simulations for different
Hurst parameters ranging from $H=0.1$ (lowest curve) to $H=0.8$ (uppermost
curve). Here, $D$ was taken to be equal to $1$, the time step $\delta t=1$,
$N_\mathrm{stat}=20,000$, and $N_\mathrm{max}=2^{10}\approx10^3$.}
\end{figure}

\begin{figure}
\includegraphics[width=8.8cm]{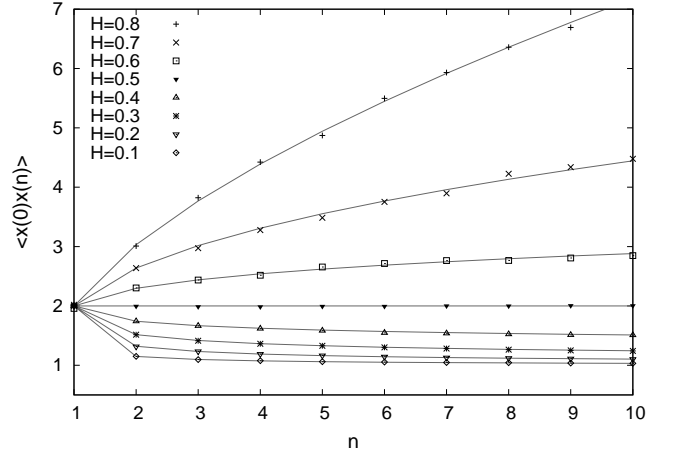}
\caption{\label{msd_acf} Position autocorelation function of free fBm. The
solid lines show the analytical expression (\ref{eq:covar-fbm-discr}) while
the symbols depict the simulations for different Hurst parameters ranging
from $H=0.1$ (lowest curve) to $H=0.8$ (uppermost curve). Again, $D$ was
taken to be equal to $1$, the time increment $\delta t=1$, $N_\mathrm{stat}=2
\times 10^6$, and $N_\mathrm{max}=64$.}
\end{figure}

Second, we calculated the position mean squared displacement
\begin{equation}
\langle x_H(t)^2\rangle=2Dt^{2H}.
\end{equation}
and two-point correlation function
\begin{equation}
\langle x_H(t_1)x_H(t_2)\rangle=D(t_1^{2H}+t_2^{2H}-|t_1-t_2|^{2H}).
\label{eq:covar-fbm}
\end{equation}
of free FBM, and compare with the analytical expressions for FBM in discrete
time $n$ with time increments $\delta t=1$,
\begin{eqnarray}
\label{msd-discr}
\langle x_H(n)^2\rangle&=&2D n^{2H},\\
\langle x_H(n)x_H(1)\rangle&=&D\left(1+n^{2H}-|n-1|^{2H}\right).
\label{eq:covar-fbm-discr}
\end{eqnarray}
As demonstrated in Figs.~\ref{msd1} and \ref{msd_acf}, respectively, the
agreement is excellent.

\begin{figure}
\includegraphics[width=8.8cm]{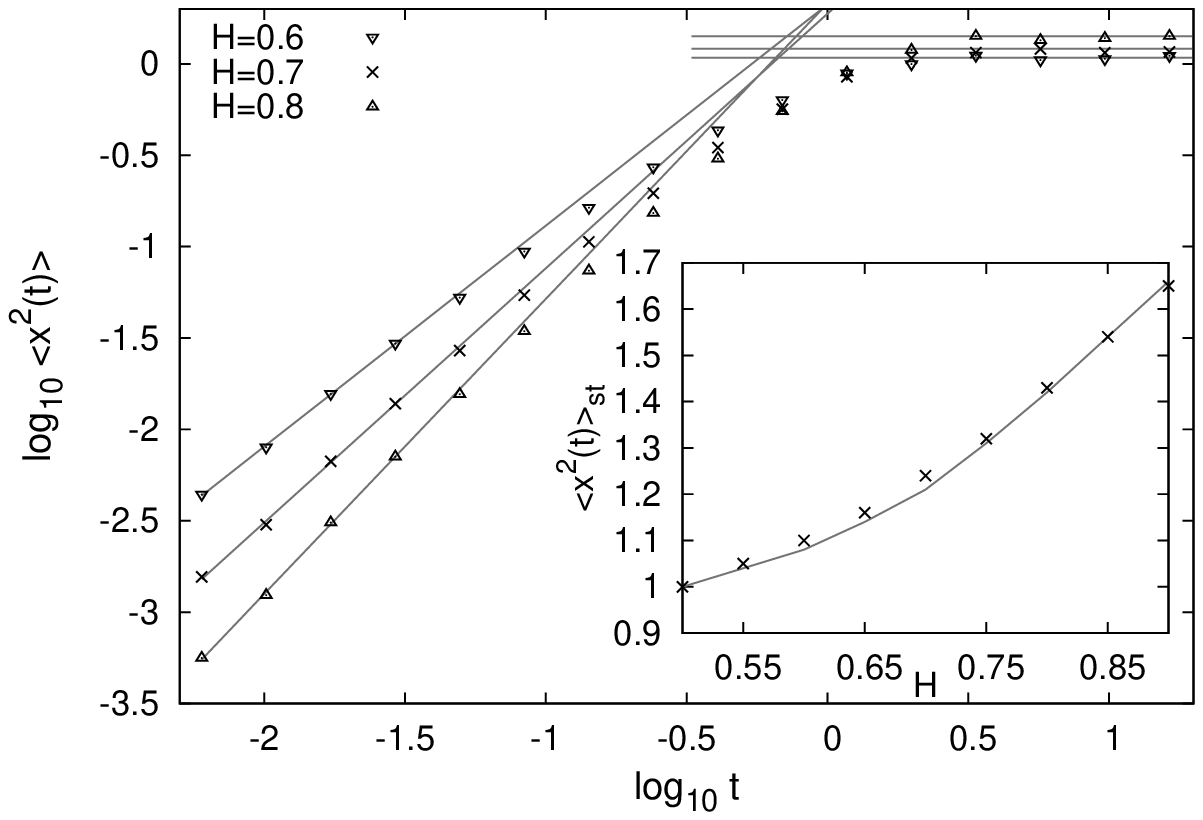}
\includegraphics[width=8.8cm]{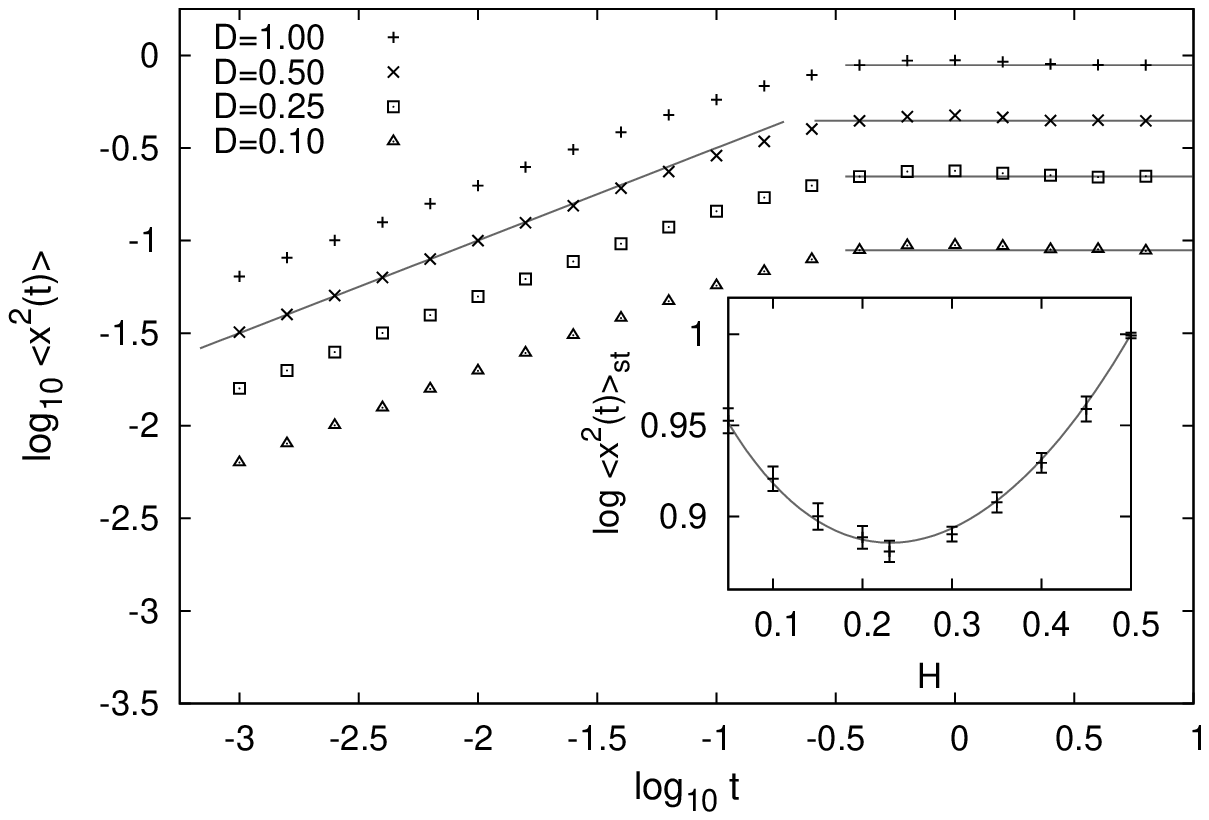}
\caption{Mean squared displacement of FBM in an harmonic potential. Top
panel: mean squared displacement for fixed $D=1.0$ and varying Hurst index.
Bottom
panel: mean squared displacement for fixed Hurst index $H=0.25$ and four
different values of the noise intensity. The solid
lines show the asymptotes of free FBM ($t^{2H}$ power-laws corresponding to
straight lines in the log-log scale) and stationary states (horizontal lines).
Insets: stationary values of the mean squared displacements as functions of
$H$ for fixed $D=1.0$. The points in all graphs represent the simulations
results for the following parameters: time increment $\delta t=0.01$,
number of samples $N_\mathrm{stat}=10^6$, and number of steps per sample
$N_\mathrm{max}=2^{10}\approx10^3$ for both persistent and antipersistent
cases.}
\label{harmonic}
\end{figure}

Third, solving Eq.~(\ref{LE_discr}) we calculated the mean squared
displacement for a particle in an infinite harmonic potential well,
as shown in Fig.~\ref{harmonic}. The initial condition was $x=0$,
at the bottom of the potential well. The asymptotic analytical
behaviors are represented by the initial free behavior $\langle
x^2(t)\rangle\simeq t^{2H}$ and the terminal saturation value
$\left\langle x^2(t)\right\rangle_\mathrm{st} =D\Gamma(1+2H)$ at
$t\rightarrow\infty$ (for details, see Appendix \ref{appendixC}).
This demonstrates that our generators also produce reliable behavior
in an external potential.

\begin{figure}
\includegraphics[width=8.8cm]{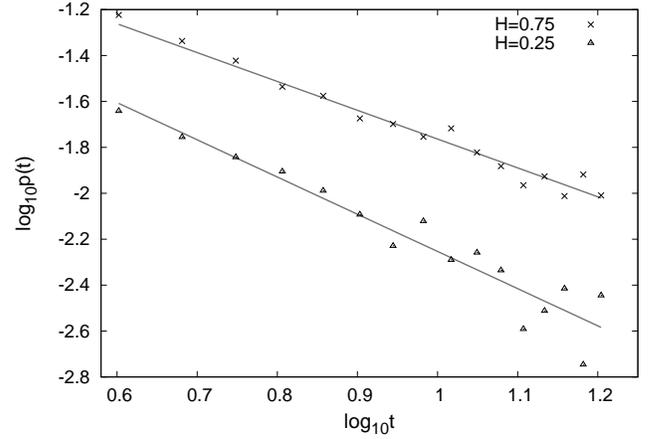}
\caption{\label{fig1} First passage time PDF of FBM on a semi-infinite axis
with absorbing boundary condition. The solid lines
demonstrate the respective analytical slopes. Parameters in the antipersistent
case: $H=0.25$, time increment $\delta t=0.001$, number of samples $N_\mathrm{
stat}=100,000$ and number of steps per sample $N_\mathrm{max}=2^{17}\approx
1.3\times10^5$. In the persistent case we used $H=0.75$, $\delta t=0.001$,
$N_\mathrm{stat}=20,000$, and $N_\mathrm{max}=2^{13}\approx8.2\times10^3$.}
\end{figure}

Finally, we performed a simulation of a free particle escaping from
a semi-infinite axis with absorbing boundary under the influence of
fractional Gaussian noise, see Fig.~\ref{fig1}. The observed scaling of
the first passage time PDF $p\left(t\right)$ compares very favourably with
the analytical result from Refs.~\cite{Molchan}:
\begin{equation}
p(t)\simeq t^{-2+H}.
\end{equation}
Note that this relation cannot be obtained by the method of images, despite
the fact that FBM is a Gaussian process. Also note that the slope of this
power-law becomes flatter for increasing Hurst coefficient: the escape is
slower for a more persistent FBM, i.e., a motion whose mean squared
displacement grows faster. This a priori surprising behavior is also seen
for the escape from the potential well studied herein.

\section{Variance and autocorrelation function for FBM in a
harmonic potential well.}
\label{appendixC}

We now consider FBM in a harmonic potential, as described by the
Langevin equation (compare with Eq.~\eqref{LEdl})
\begin{equation}
\label{eq_3_1_} \frac{dx(t)}{dt}=-a x+D^{1/2}\xi_H(t),
\end{equation}
where we introduce the prefactor $a$ which allows us to consider the
harmonic potential ($a=1$) and a free FBM ($a=0$) as well. The
solution of Eq.~\eqref{eq_3_1_} with the initial condition
$x(t=0)=0$ is
\begin{equation}
\label{eq_3_2_} x(t)=D^{1/2}\int_0^te^{-a (t-t')}\xi_H(t')dt'.
\end{equation}
Then, the ACF function
\begin{eqnarray}
\nonumber \left\langle x(t_1)x(t_2) \right\rangle  =
 De^{ -2 a t} \int\limits_0^{t_1} {dt' \int\limits_0^{t_2} {dt''e^{a (t'+t'')} \left\langle \xi(t')\xi(t'') \right\rangle}}   \\
= - De^{ - 2 a t} \int\limits_0^{t_1} {dt' \int\limits_0^{t_2}
{dt''e^{a (t'+t'')} \frac{{\partial ^2 }}{{\partial t'\partial t''}}\left|
{t' - t''} \right|^{2H} } }.
\end{eqnarray}
Now, if $t_2-t_1=\tau$, $\tau > 0$, after some lengthy calculations we get Eq.~(\ref{ACFfull}):
\begin{widetext}
\begin{eqnarray}
\label{ACFfull}
\nonumber \left\langle x(t) x(t+\tau) \right\rangle &=&
D \Bigg \{ e^{-a (t+\tau)} t^{2 H} + e^{-a t} (t+\tau)^{2 H}- \tau^{2 H}\\
\nonumber &-& \frac{2 a^2-1}{2a(2 H +1)} \Big [ t^{2 H+1}
e^{-a (2 t+\tau)} M(2 H+1;2 H+2;a t) \\
&+& (t+\tau)^{2 H+1} e^{-a(2t+\tau)}
M(2 H+1;2 H+2;a (t+\tau))\\
\nonumber &-&\ \tau^{2 H+1} e^{-a \tau} M(2 H+1;2 H+2;a \tau)\Big ]\\
\nonumber &-&\frac{1}{2} a^{-2 (H+1)} \left(2 a^2-1\right)
\Big [ e^{a \tau} (\Gamma (2 H+1;a (t+\tau))-
\Gamma (2 H+1;a \tau))  \\
\nonumber &+& e^{-a \tau} (\Gamma (2 H+1;a t)-
\Gamma (2 H+1))\Big ] \Bigg \}.
\end{eqnarray}
Assuming $a=1$,
\begin{eqnarray}
\label{ACFfull_a1}
\nonumber \left\langle x(t)x(t+\tau) \right\rangle &=& D \Bigg \{ e^{-(t+\tau)}t^{2H}- \tau^{2H}+e^{-t}(t+\tau)^{2H} \\
&+& \frac{1}{2}e^{-\tau} \left[ \Gamma(2H+1) -\Gamma(2H+1;t) + \frac{\tau^{2H+1}}{2H+1} M(2H+1;2H+2;\tau)\right] \\
\nonumber &-& \frac{1}{2}e^{-2t-\tau} \left[ \frac{t^{2H+1}}{2H+1} M(2H+1;2H+2;t)+ \frac{(t+\tau)^{2H+1}}{2H+1} M(2H+1;2H+2;t+\tau)\right] \\
\nonumber &+& \frac{1}{2}e^{\tau} \left[ \Gamma(2H+1;\tau) -\Gamma(2H+1;t+\tau) \right] \Bigg \}.
\end{eqnarray}
\end{widetext}

Here, $\Gamma(a,b)$ is the incomplete $\Gamma$-function, and $M$
denotes the Kummer function \cite{Abramowitz}. In the stationary
state ($t\to\infty$) the autocorrelation function
Eq.~(\ref{ACFfull_a1}) yields
\begin{eqnarray}
\nonumber &&\left\langle
x(t)x(t+\tau)\right\rangle_{st}= D \Bigg \{\left[e^{-\tau} \Gamma(2H+1)+e^{\tau}\Gamma(2H+1,\tau) \right]\\
&&+ \frac{\tau^{2H+1}e^{-\tau}}{2(2H+1)}M(2H+1;2H+2,\tau) -\tau^{2H} \Bigg\}.
\label{eq_3_5_}
\end{eqnarray}

In order to obtain the variance we take $\tau=0$ in Eq.~(\ref{ACFfull_a1}):
\begin{eqnarray*}
\left\langle {x^2 \left( t \right)} \right\rangle  &=& D \Bigg \{ 2t^{2H} e^{-t}  + \left[ \Gamma \left( 1 + 2H \right) - \Gamma \left( 1 + 2H,t \right) \right] \\
  &-&  \frac{t^{2H+1}}{2H+1}e^{-2t} M(2H+1;2H+2;t) \Bigg \}.
\end{eqnarray*}
Now, the stationary variance is:
\begin{equation}
\label{eq_3_4_} \left\langle x^2\right\rangle_{st}=D\Gamma(2H+1).
\end{equation}

Note that at $\tau=0$ Eq.~\eqref{eq_3_5_} reduces to
Eq.~\eqref{eq_3_4_}, whereas for $H = 1/2$ it gives $\left\langle
x(t)x(t+\tau)\right\rangle_{st} =De^{-\tau}$, the autocorrelation
function of the Ornstein-Uhlenbeck process. Taking the asymptotics of the incomplete $\Gamma$-function and the Kummer function, one may easily
see that $\left\langle
x(t)x(t+\tau)\right\rangle_{st}\approx2DH(2H-1)\tau ^{2H-2}$ at
$\tau\to\infty$.

The autocorrelation function of free fBm can be naturally obtained by placing $a=0$ in Eq.~(\ref{ACFfull}):
\begin{equation}
\left\langle x(t) x(t+\tau) \right\rangle =
D \left\{  t^{2 h}+
    (t+\tau)^{2 h}-
   \tau^{2 h}\right\},
\end{equation}
\noindent that matches the well-known relation \cite{Qian}.

\section{Mean escape time and first escape time PDF for harmonic potential truncated
from both sides}
\label{appendixD}

\begin{figure}
\includegraphics[width=8.8cm]{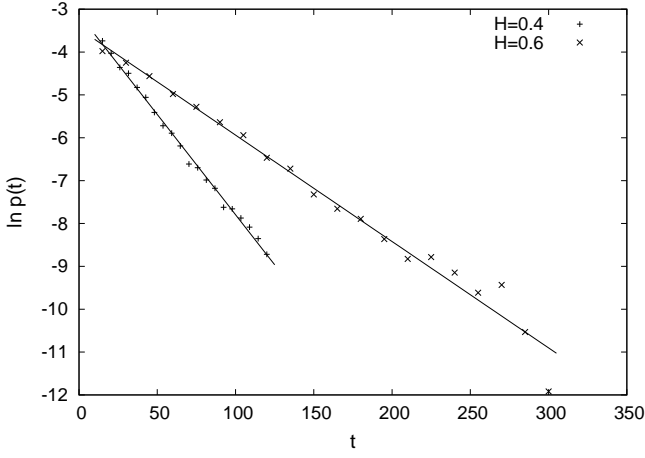}
\caption{\label{fig_a} First escape time PDF for harmonic potential
truncated from both sides. Points are the simulation data, solid
lines stand for linear fitting. Simulation details are the
following: for the antipersistent case $D=0.25$, $\delta t=0.001$,
$N_\mathrm{max}=131072$, $N_\mathrm{stat}=10^5$; for the persistent
case $D=0.25,  \delta t=0.002, N_\mathrm{max}=131072,
N_\mathrm{stat}=10^5$.}
\end{figure}

\begin{figure}
\includegraphics[width=8.8cm]{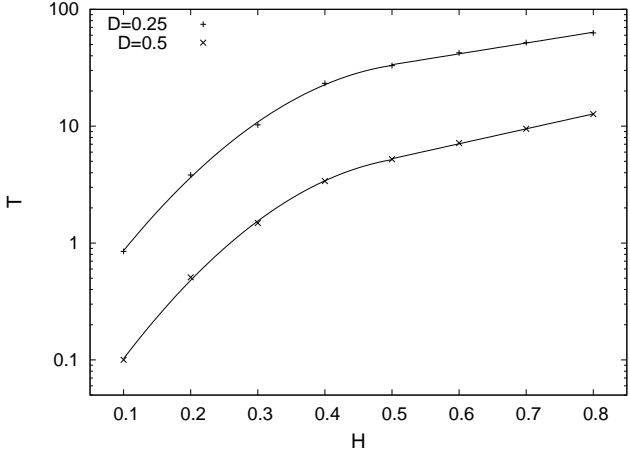}
\caption{\label{fig_b} Mean escape time as function of the Hurst
parameter for harmonic potential truncated from both sides. Points
are the simulation data, solid lines stand for fitting with
Eq.~(\ref{fit_2tr}). Simulation details are the following: for both
antipersistent and persistent cases $\delta t$ varied from $0.001$
to $0.005$, $N_\mathrm{stat}=10^5$,
$N_\mathrm{max}=2^{13}\ldots2^{21}\approx8\times10^3\ldots2\times10^6$. }
\end{figure}

In this Appendix we consider the Kramers problem for an harmonic potential,
but this time we introduce a cutoff on both sides, that is, at
$x=\pm \sqrt{2}$, and evaluate the same dependencies (see Figures
\ref{fig_a} and \ref{fig_b}).

One can observe that qualitatively there is no difference in
behaviour with the case of the one-side truncated potential. Indeed,
the escape is faster when lowering the Hurst parameter; the escape
time PDF remains exponential and so does the mean escape time.
Again, the MET may be fitted with the following function:

\begin{eqnarray}
\label{fit_2tr}
\nonumber T(H \leq 0.5) &=& \exp(ax^2+bx+c) \\
T(H > 0.5) &=& \exp(b'x+c'),
\end{eqnarray}
where $a, b, c, b', c'$ are some constants depending on $D$.

\end{appendix}

\newpage


\begin{thebibliography}{99}

\bibitem{sinai} Y. Sinai, Theor. Prob. Appl. \textbf{27}, 256 (1982).

\bibitem{julia} J. Dr{\"a}ger and J. Klafter, Phys. Rev. Lett. \textbf{84},
5998 (2000).

\bibitem{vincent1} V. Tejedor and R. Metzler, J. Phys. A \textbf{43},
082002 (2010).

\bibitem{richardson} L. F. Richardson, Proc. Roy. Soc. London A \textbf{110},
709 (1926).

\bibitem{boffetta} G. Boffetta and I. M. Sokolov, Phys. Rev. Lett. \textbf{88},
094501 (2002).

\bibitem{bouchaud} J.-P. Bouchaud and A. Georges, Phys. Rep. \textbf{195},
127 (1990).

\bibitem{report} R. Metzler and J. Klafter, Phys. Rep. \textbf{339}, 1 (2000);
J. Phys. A \textbf{37}, R161 (2004).

\bibitem{scher} H. Scher and E. W. Montroll, Phys. Rev. B \textbf{12}, 2455
(1975);

\bibitem{pfister} G. Pfister and H. Scher, Adv. Phys. \textbf{27}, 747 (1978);
Q. Gu, E. A. Schiff, S. Grebner, and R. Schwartz, Phys. Rev. Lett. \textbf{76},
3196 (1996).

\bibitem{scher_grl} H. Scher, G. Margolin, R. Metzler, J. Klafter, and B.
Berkowitz, Geophys. Res. Lett. \textbf{29}, 1061 (2002);
B. Berkowitz, A. Cortis, M. Dentz and H. Scher, Reviews of Geophysics, 44,
RG2003 (2006).

\bibitem{klemm} A. Klemm, R. Metzler, and R. Kimmich, Phys. Rev. E \textbf{65},
021112 (2002); S. Havlin and D. ben-Avraham, Adv. Phys. \textbf{36}, 695
(1987).

\bibitem{bio} A. Caspi, R. Granek, and M. Elbaum, Phys. Rev. Lett.
\textbf{85}, 5655 (2000);
I. M. Toli{\'c}-N{\o}rrelykke et al., \emph{ibid.} \textbf{93}, 078102 (2004);
I. Golding and E. C. Cox, \emph{ibid.} {\bf 96}, 098102 (2006);
H. Yang et al., Science \textbf{302}, 262 (2003);
M. Weiss, M. Elsner, F. Kartberg, and T. Nilsson, Biophys. J. \textbf{87},
3518 (2004);
G. Seisenberger et al., Science \textbf{294}, 1929 (2001).

\bibitem{alabio} I. Y. Wong et al., Phys. Rev. Lett. \textbf{92}, 178101(2004);
W. Pan et al., \emph{ibid.} {\bf 102}, 058101 (2009);
D. Banks and C. Fradin, \emph{ibid.} {\bf 89}, 2960 (2005).

\bibitem{caspi} A. Caspi, R. Granek, and M. Elbaum, Phys. Rev. E \textbf{66},
011916 (2002).

\bibitem{matheron} G. Matheron and G. de Marsily, Water Res. Res. \textbf{16},
901 (1980).

\bibitem{swinney} T. H. Solomon, E. R. Weeks, and H. L. Swinney, Phys. Rev.
Lett. \textbf{71}, 3975 (1993).

\bibitem{stapf} S. Stapf, R. Kimmich, and R.-O. Seitter, Phys. Rev. Lett.
\textbf{75}, 2855 (1995); O. V. Bychuk and B. O'Shaugnessy, J. Chem. Phys.
\textbf{101}, 772 (1994); A. V. Chechkin, I. M. Zaid, M. A. Lomholt, I. M.
Sokolov, and R. Metzler, Phys. Rev. E \textbf{79}, 040105(R) (2009).

\bibitem{klablushle} J. Klafter, A. Blumen and M. F. Shlesinger,
Phys. Rev. A {\bf 35}, 3081 (1987).

\bibitem{mekla} R. Metzler and J. Klafter, Chem. Phys. Lett. \textbf{321},
238 (2000).

\bibitem{ditlevsen} P. D. Ditlevsen, Phys. Rev. E \textbf{60}, 172 (1999).

\bibitem{levykramers} A. V. Chechkin, V. Yu. Gonchar, J. Klafter, and R.
Metzler, Europhys. Lett. \textbf{72}, 348 (2005); A. V. Chechkin, O. Yu.
Sliusarenko, J. Klafter, and R. Metzler, Phys. Rev. E. \textbf{75}, 041101
(2007).

\bibitem{imkeller} P. Imkeller and I. Pavlyukevich, J. Phys. A \textbf{39},
L237 (2006).

\bibitem{kolmogorov} A. N. Kolmogorov, Dokl. Acad. Sci. USSR \textbf{26},
115 (1940).

\bibitem{mandelbrot} B. B. Mandelbrot and J. W. van Ness, SIAM Rev. \textbf{1},
422 (1968). Compare also
B. B. Mandelbrot, Physica Scripta \textbf{32}, 257 (1985).

\bibitem{Yaglom} A. Yaglom, Correlation theory of stationary and
related random functions (Springer, Berlin, 1987).

\bibitem{Qian} H. Qian, Fractional Brownian Motion and Fractional
Gaussian Noise. In G. Rangarajan and M.Z. Ding (eds), {\it Processes
with Long-Range Correlations} (Springer, Lecture Notes in Physics,
Vol.621), pp.22-33.

\bibitem{gleb} D. Panja, E-print arXiv:0912.2331.

\bibitem{tobias} L. Lizana and T. Ambj{\"o}rnsson, Phys. Rev. Lett.
\textbf{100}, 200601 (2008); Phys. Rev. E \textbf{80}, 051103 (2009).

\bibitem{guigas} G. Guigas and M. Weiss, Biophys. J. \textbf{94}, 90 (2008);
J. Szymanski and M. Weiss, Phys. Rev. Lett. \textbf{103}, 038102 (2009);
V. Tejedor et al, Biophys J. (at press).

\bibitem{hurst} H. E. Hurst, Trans. Amer. Soc. Civil Eng. \textbf{116}, 400
(1951).

\bibitem{palmer} T. N. Palmer, G. J. Shutts, R. Hagedorn, F. J. Doblas-Reyes,
T. Jung, and M. Leutbecher, Ann. Rev. Earth Planet. Sci. \textbf{33}, 163
(2005).

\bibitem{mandelbrot1} I. Simonsen, Physica A \textbf{322}, 597 (2003);
N. E. Frangos, S. D. Vrontos, and A. N. Yannacopoulos, Appl. Stochast.
Models in Business and Industry \textbf{23}, 403 (2007).

\bibitem{mikosch} T. Mikosch, S. Rednick, H. Rootz{\'e}n, and A. Stegemann,
Ann. Appl. Prob. \textbf{12}, 23 (2002).

\bibitem{Molchan} M. Ding and W. Yang, Phys. Rev. E \textbf {52}, 207 (1995);
J. Krug et al. Phys. Rev. E \textbf{56}, 2702 (1997); G.M. Molchan.
Commun. Math. Phys. {\bf 205} 97 (1999).

\bibitem{stas} S. Burov and E. Barkai, Phys. Rev. Lett. \textbf{100}, 070601
(2008).

\bibitem{eli} E. Barkai and R. Silbey, Phys. Rev. Lett. \textbf{102},
050602 (2009).

\bibitem{katja} A. Romero, J. M. Sancho, and K. Lindenberg, Fluct. and Noise
Lett. \textbf{2}, L79 (2002).

\bibitem{goychuk} I. Goychuk and P. H{\"a}nggi, Phys. Rev. Lett. \textbf{99},
200601 (2007); compare also I. Goychuk E-print arXiv:0905.082.

\bibitem{klimontovich} Yu. L. Klimontovich, Turbulent motion and the
structure of chaos: a new approach to the statistical theory of open systems
(Kluwer, Dordrecht, The Netherlands, 1992).

\bibitem{kramers} H. A. Kramers, Physica A \textbf{7}, 284 (1940).

\bibitem{chandrasekhar} S. Chandrasekhar, Rev. Mod. Phys. {\bf 15} 1 (1943).

\bibitem{risken} H. Risken, The Fokker-Planck equation (Springer-Verlag,
Berlin, 1989).

\bibitem{fGnAlg1}  B.S. Lowen Methodology and Computing in Applied Probability
{\bf 1:4}, 445 (1999).

\bibitem{fGnAlg2} A.V. Chechkin and V.Yu. Gonchar, Chaos, Solitons and Fractals
\textbf{12}, 391 (2000).

\bibitem {Slepian} D. Slepian, Bell Syst. Tech. J. \textbf{41}, 463 (1962).

\bibitem{Rice} S. O. Rice, Bell Syst. Tech. J. \textbf{23}, 282 (1944);
\emph{ibid.} \textbf{24}, 46 (1945), reproduced in Noise and Stochastic
Processes, edited by N. Wax (Dover, New York, NY, 1954).

\bibitem{stratonovich} R. L. Stratonovich, Topics in the theory if random
noise, Vol.~II (Gordon and Breach, New York, NY, 1967).

\bibitem{haenggi} P. H{\"a}nggi and P. Jung, Adv. Chem. Phys. \textbf{89}, 239
(1995).

\bibitem{willemski} G. Wilemski and M. Fixman, J. Chem. Phys. \textbf{60}, 866
(1974); \emph{ibid.}, 878 (1974).

\bibitem {szabo} A. Szabo, K. Schulten, Z. Schulten, J. Chem. Phys.
\textbf{72}, 4350 (1980).

\bibitem {Igor-Tatiana} T. Verechtchaguina, I.M. Sokolov, and L.
Schimansky-Geier, Phys. Rev. E \textbf{73}, 031108 (2006).

\bibitem {Igor_PRL} I. M. Sokolov, Phys. Rev. Lett. \textbf{90},
080601 (2003).

\bibitem{redner} S. Redner, A guide to first passage processes (Cambridge
University Press, Cambridge, UK, 2001).

\bibitem{hughes} B. D. Hughes, Random walks and random environments. Vol. 1:
Random Walks (Clarendon Press, Oxford, UK, 1995). Cf. chapter 3.2.

\bibitem {Marques} A.E. Likthman, C.M. Marques, Europhys. Lett.
\textbf{75}, 971 (2006).

\bibitem{deng} W. H. Deng and E. Barkai, Phys. Rev. E \textbf{79}, 011112
(2009); J.-H. Jeon and R. Metzler, Phys. Rev. E (at press).

\bibitem{web} A. Lubelski, I. M. Sokolov, and J. Klafter, Phys. Rev. Lett.
\textbf{100}, 250602 (2008); Y. He, S. Burov, R. Metzler, and E. Barkai,
\emph{ibid.} \textbf{101}, 058101 (2008); R. Metzler, V. Tejedor, J.-H. Jeon,
Y. He, W. Deng, S. Burov, and E. Barkai, Acta Phys. Polonica B \textbf{40},
1315 (2009); T. Neusius, I. M. Sokolov, and J. C. Smith, Phys. Rev. E
\textbf{80}, 011109 (2009); S. Burov, R. Metzler, and E. Barkai (unpublished).

\bibitem{Abramowitz} M. Abramowitz, I.A. Stegun, Handbook of Mathematical Functions
(National Bureau of Standards, Tenth Printing, USA, 1972).

\end{thebibliography}
\end{document}